# USE OF A NIGHT-TRACKING CAMERA FOR CHARACTERIZATION AND ORBIT IMPROVEMENT OF DEFUNCT SPACECRAFT

**E. Cordelli\*, P. Schlatter, P. Lauber, T. Schildknecht**

*Astronomical Institute University of Bern, Sidlerstrasse 5, CH-3012 Bern, Switzerland*
*\*Corresponding author, email: emiliano.cordelli@aiub.unibe.ch*

**ABSTRACT**

The precise knowledge of the positions of space debris objects and in particular of defunct satellites is fundamental for satellite operations. Several studies showed that it is possible to improve the accuracy of the orbit determination results by fusing different types of observables, i.e. classical astrometric positions and range measurements. Particularly promising in the space debris field are the ranges provided by a satellite laser ranging system. The factors that limit the applicability of the satellite laser ranging (SLR) techniques are the altitude of the target, the accuracy of the predicted ephemeris of the target, the energy of the laser pulse, and the laser field of view.

In this paper we will show a way to overcome the mentioned challenges by using a night-tracking camera for the real time correction of the pointing of the SLR system (active tracking), and for the simultaneous acquisition of measurements used to improve the orbits and to study the attitude of the target. After presenting the basic functionalities, the performance of the night-tracking camera, and the procedure to acquire the measurements, we will also show the potential of this tool to allow improving orbits in real-time. This study is carried out for defunct or recently decommissioned satellites. Only real angular/laser measurements provided by the sensors of the Swiss Optical Ground Station and Geodynamics Observatory Zimmerwald (SwissOGS) owned by the Astronomical Institute of the University of Bern (AIUB) are used.

## 1  INTRODUCTION

The space debris constitutes a serious issue for today's space activities. In order to ensure the usability of the outer space, some actions are needed. The two main approaches are: the prevention of the creation of new space debris, and their removal from space. Excluding the mitigation techniques which deal with the secure disposal of space objects after their life time, the collision avoidance maneuvers and the active debris removal operations require the precise knowledge of the orbit and of the attitude of a space debris object.

Studies [1, 2] have shown that particularly promising in the improvement of the orbit determination (OD) accuracy, especially when processing short observation arcs, is the fusion of range and angular data. One way to obtain range data is via a laser system. Furthermore, the laser range can also be used for the attitude determination of defunct satellites as shown in [3, 4]. The main limiting factors of the satellite laser ranging (SLR) technique for space debris application are: the accuracy of the predicted orbit and the field of view (FoV) of the laser. One way to overcome these problems is to use a tracking camera.

In this paper, we will show the characteristics and the capability of the tracking camera implemented recently at the SwissOGS. We will then focus on the output data provided by the camera and how this can be used. Finally, we will show some OD results obtained for LEO and MEO satellites.

## 2  CAMERA INTEGRATION & CAPABILITY

The integration of the tracking camera was performed on the ZIMLAT telescope which hosts also the SLR system of the SwissOGS. For more details in the camera integration please refer to [5]. The camera and the laser system have to work concurrently. For that reason, since the laser system is in the Coudé focus of the telescope, the camera was installed in the Nasmyth focus. The effective focal length of the chosen focal station is 8m. This provides a relatively small FoV (~7 arcmin) but it is optimal for blocking with a notch filter the laser light. A sketch of the telescope architecture where it is possible to see the laser path and the camera position is reported in Figure 1.

The camera needed to be integrated also on the software level. Two main software integrations were carried out. First, we modified the SLR system to make it able to communicate with the camera, then we developed the software for the camera control. This software, whose screenshot is visible in Figure 2, is used to set up the camera parameters as: exposure time, readout mode and speed, and the binning. However, the main aim of the software is the calculation of the offset of the position of the object w.r.t. the position of the laser on to the camera chip. The laser pointing direction on the image is indicated by the red cross of Figure 2. The determined ephemeris offset is sent to the telescope as pointing correction. Therefore, the software can communicate with and monitor the SLR system. The last feature of the tracking camera software is the capability of storing images that contain the brightness information of the object and, in their header, the telescope parameters and the precise measurement epoch provided by the SLR system.

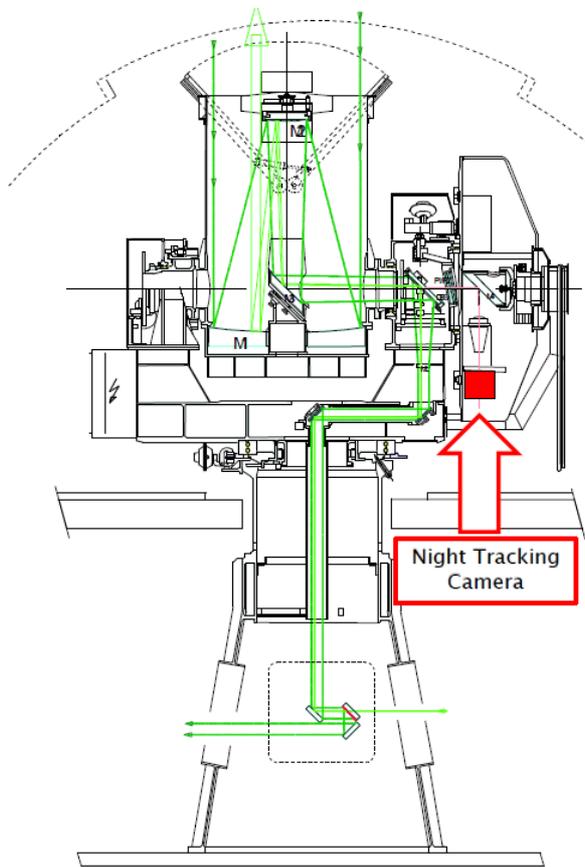

*Figure 1 Installation of the night-tracking camera on ZIMLAT telescope.*

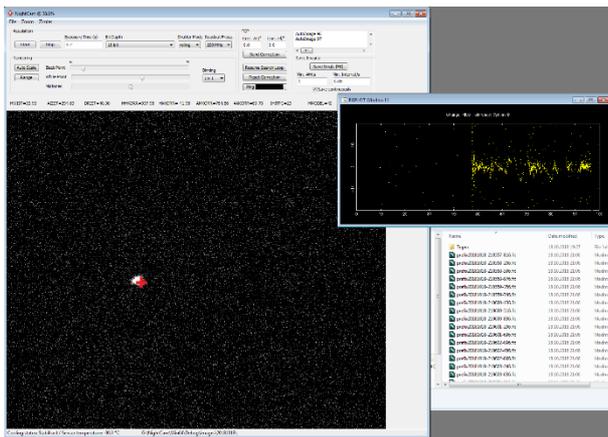

*Figure 2 Screenshot of the night-tracking camera software.*

## 3  CAMERA OUTPUT

The direct outputs of the tracking camera are the angular positions of the object in the sky, and its brightness extracted from the images. Once the pointing of the telescopes has been corrected the SLR system provides also the distance of the object.

The camera, being able to correct the telescope pointing in real time (and therefore the a priori ephemeris), provides the actual azimuth and elevation of the object in the sky without the need of an astrometric data reduction procedure. In fact, the azimuth and elevation positions are provided directly by the angular encoder of the telescope. This result is quite useful especially for LEO objects where, due to the small FoV and the short exposure times, usually not enough stars are captured in the image. The main disadvantage is a lower accuracy of the angular measurements w.r.t. the classical astrometric data reduction since the accuracy it is now dependent on the mount model accuracy of the telescope. Nevertheless as we will see in paragraph 5, these angular positions are fundamental for this special type of OD, e.g. on-the-fly OD.

Once the pointing direction of the telescope has been corrected with the help of the camera, we are able to measure the distance to the object using the SLR system. The combination of the two observables is used to perform an OD almost in real time, allowing us to update the ephemeris and re-observe a LEO object in the next pass.

The ranges, together with the brightness information extracted from the images stored by the camera, can be used for the attitude determination of the object.

We would like to highlight two aspects. The first is the measurement frequency; the SLR system is working at ~100Hz pulse rate while the sCMOS camera allows us to acquire high-resolution light curves (up to 30 full frames per second). The second aspect is given by the fact that all the measurements are acquired at the same time, this will allow us to perform both the orbit determination and the analysis of the attitude status of the object within one observation session.

Figure 3 and Figure 4 show the observations, resulting from a pass of ENVISAT and TOPEX, acquired by the tracking camera. The graph on the right shows the angular position of the object in the sky, while the first two on the left show the measured distance and the detrended one, in terms of difference between expected and measured time of flight. The detrended is reported to highlight the changes in ranges due to attitude motion of the satellite. Finally, the last plot shows the light curve extracted from the stored images. The interruptions in the ENVISAT light curve are caused by an option of the tracking camera software which was storing images only when the telescope was able to range to the object. Looking at the 2nd and 3rd left-plot of Figure 4, one can appreciate the high resolution of the light curve whose periodicity is consistent with that visible for the ranges.

*Figure 5* shows the results obtained by the tracking camera for a spent Glonass (cospar ID: 91025B). The light curve periodicity is coincident with the accepted ranges by the SLR system (yellow dots). The shape of the accepted ranges shows the movement of the retroreflector around the center of mass of the satellite. This periodicity in the ranges helps to identify which portion of the light curve is produced by the nadir face of the satellite.

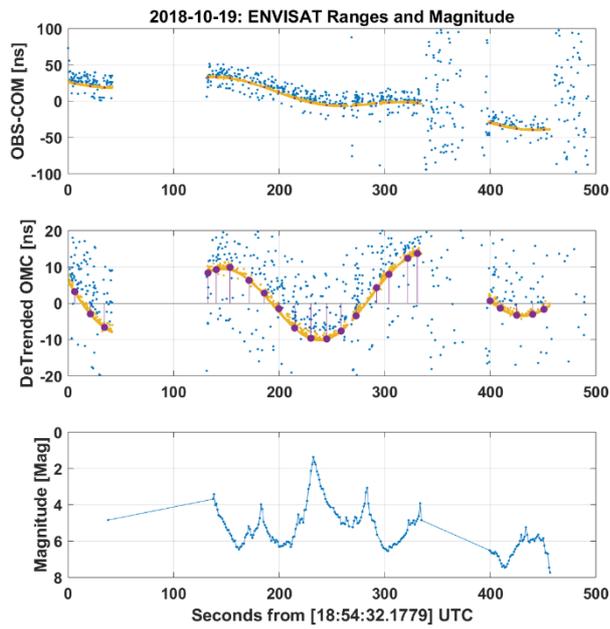 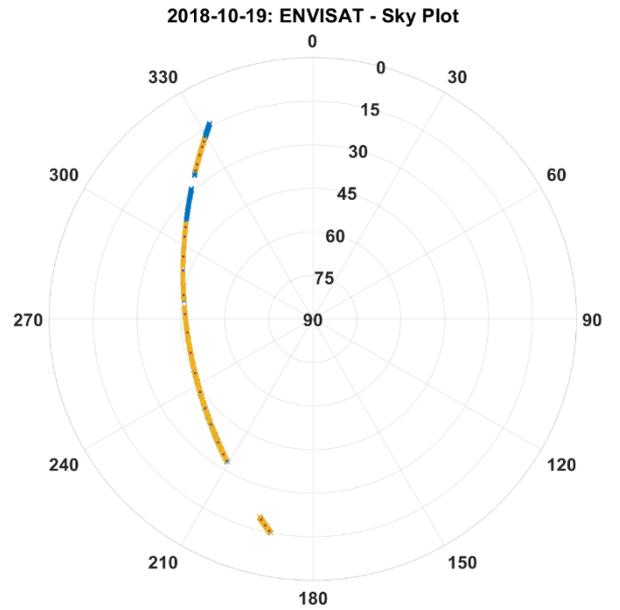

*Figure 3 Example of tracking camera output: ENVISAT.*

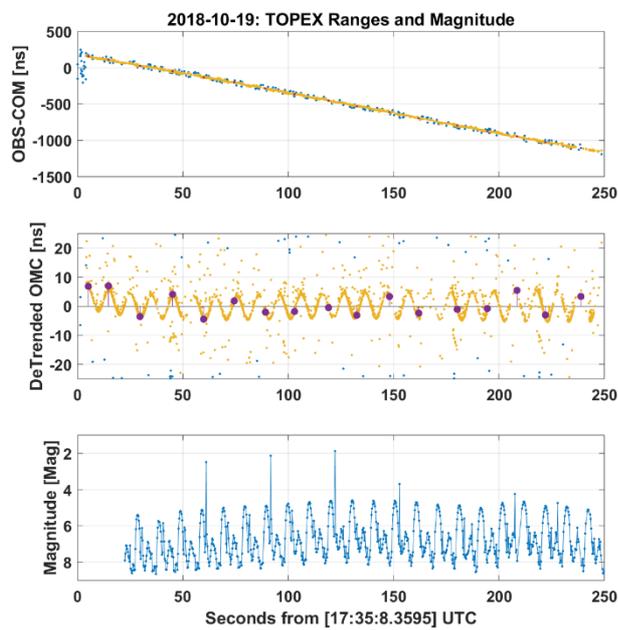 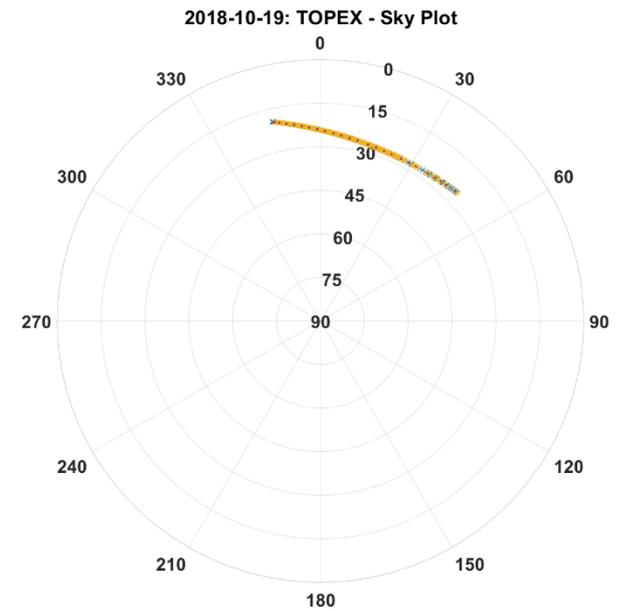

*Figure 4 Example of tracking camera output: TOPEX.*

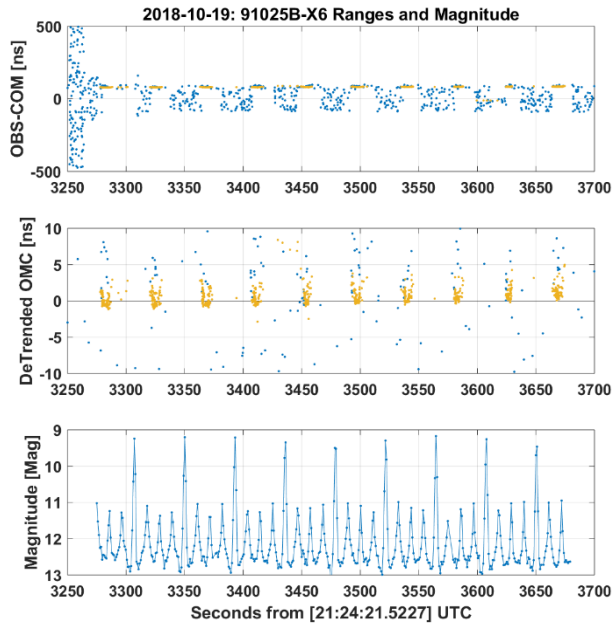
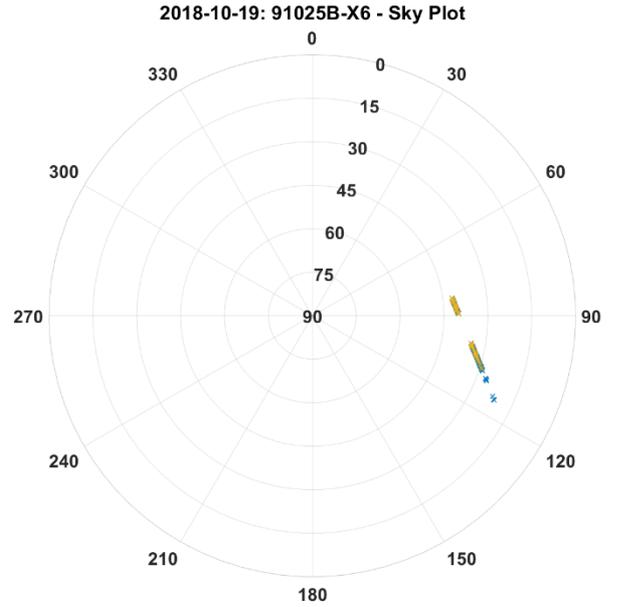

*Figure 5 Example of tracking camera output: GLONASS.*

## 4 ATTITUDE DETERMINATION RESULTS

The data collected by the night-tracking camera can be used both, for attitude and orbit determination (see Paragraph 5). The easiest parameter that can be estimated from the light curves is the spin period of the objects. Several studies are carried out at the AIUB to exploit the light curves and the range variations due to the attitude motion to determine the spin axis direction, and to study the effects of the space environmental torques [6, 7].

We applied the phase reconstruction method [8], to process the light curves acquired by the night-tracking camera and extract the spin period of the object. Several light curves acquired for TOPEX were processed. The obtained rotation periods are summarized in Table 1. The values were compared and confirmed by those present in the AIUB database [9]. The same procedure was applied to the data acquired for the Glonass 91025B. Two distinct observation sessions were performed during the same night, for both data sets the extracted apparent rotation period is ~42,925 ± 0.05 sec. This value is also confirmed by the results contained in our database [9].

*Table 1 Summary of extracted apparent rotation periods for TOPEX.*

| Date | | Extracted Rotation Period |
|---|---|---|
| 2018-09-25 | | 10.235 ± 0.01 sec |
| 2018-10-09 | | 10.207 ± 0.002 sec |
| 2018-10-17 | 1st Pass | 10.188 ± 0.004 sec |
| | 2nd Pass | 10.246 ± 0.002 sec |
| 2018-10-19 | 1st Pass | 10.189 ± 0.002 sec |
| | 2nd Pass | 10.215 ± 0.0025 sec |

The just determined apparent spin period, together with the reconstructed phase of the object, can be used to display, as shown in Figure 6, Figure 7, and Figure 8, the light curve over the pass. The high-sampling data rate of the night-tracking camera provides useful information for the object characterization. From Figure 6, being TOPEX a LEO satellite, it is possible to see how, at the beginning of the pass, due to a bigger phase angle, the difference between faintest and brightest point of the light curve is smaller than that obtained at the end of the observation series; when the satellite was entering in the shadow of the Earth. Another interesting effect that can be seen is the contribution of the two specular reflections (sharp maxima in the middle of the data series) occurring only for a limited period of the pass. Due to the Glonass angular velocity in the sky (~30 arcsec/sec), the change of the observation geometry obtained during one single observation session does not produce appreciable changes in the light curve. Looking at Figure 7 and Figure 8 independently, one can see how the brightness values are almost constant during each observation session. On the other hand, comparing these last two figures, one can see how, in Figure 7, there is a smaller difference between relative and absolute maxima of the light curve and the average brightness of the object is about 11.5 Mag.. Figure 8, instead, shows smaller brightness values but an increase of the brightness difference between relative and absolute maxima of the light curve. This is probably due to the interaction between observation-geometry and spin axis direction. In particular, on one hand the phase angle is increasing, on the other we suppose that the rotation axis of the satellite is oriented in a way that one portion of the satellite reflects more directly the sunlight to the observer.

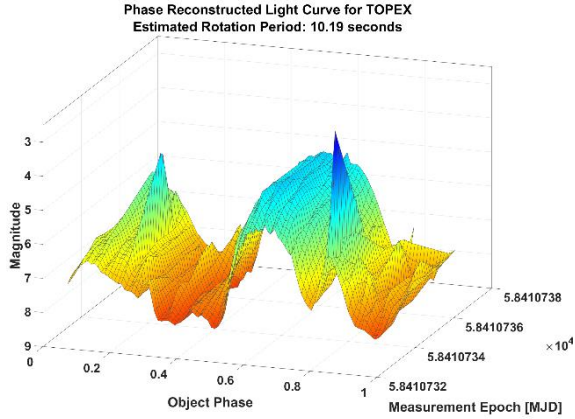

*Figure 6 Phase reconstructed light curve for one pass of TOPEX.*

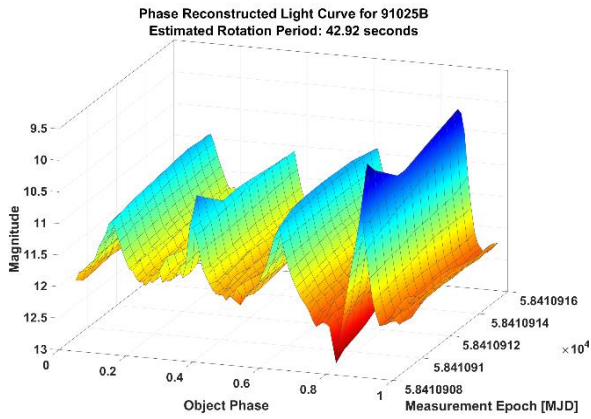

*Figure 7 Phase reconstructed light curve for Glonass 91025B (1st data set).*

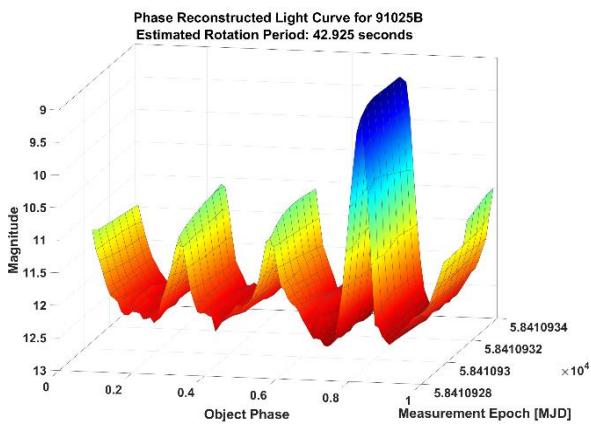

*Figure 8 Phase reconstructed light curve for Glonass 91025B (2nd data set).*

## 5  ORBIT DETERMINATION RESULTS

The tracking camera performs an orbit improvement in real time correcting the ephemeris offset of the target object. At the same time, once locked to the object, the SLR system provides the range to the object. The angular data, provided directly by the camera without any further astrometric data reduction process, and the ranges, provided by the SLR system, are used to perform an OD of the observed object. An automatic pipeline is in development to perform the OD just at the end of the measurement acquisition, in order to generate new ephemeris that will allow the re-observation of the object during its next pass.

Before employing the data gathered from the tracking camera for OD, we needed to validate them. In particular, there was no need to validate the range data. The ranges are acquired by the SLR system, which works independently from the tracking camera and it is compliant with the International Laser Ranging Service (ILRS) requirements [10]. On the other hand, we needed to validate and evaluate the accuracy of the angular measurements. For this validation, we compared the measured azimuth and elevation positions with those calculated from the ephemeris. This test was repeated on different geodetic satellites whose precise ephemeris are provided by the ILRS. The average discrepancy between measured and given position is within a radius of 15 arcsec. This value can be improved considering the wobble of the laser pointing position on the camera depending on the telescope pointing direction (see [5] for details). The 15 arcsec are also used to calculate the weight of the angular measurements in the least square adjustment, core of the OD process [2].

Before showing the OD results obtained for spent satellites, it must be said that we validated also the performance of the OD procedure exploiting ILRS target satellites. The validation was performed via the comparison of the ephemeris generated after an OD with those provided by the ILRS [10]. For details about the angular measurements and OD validation please refer to [11].

A complete test of the camera capabilities was performed on two spent satellites TOPEX (92052A) and a Glonass (91025B), respectively. Both satellites carry retroreflectors, which make easier the tracking with the SLR system. The first belongs to the LEO orbital regime while the second is a MEO. Both satellites had poor ephemeris so, we first used the camera to correct the telescope pointing then, we could acquire the measurements used to determine their rotation period (as shown in paragraph 4) and then to determine their orbit. An extract of the measurements set is reported in Figure 4 and *Figure 5*, respectively. Since for these objects there were no reference ephemeris which we could have used to validate our OD results, we decided to use two sets of data for the OD quality check. In particular, the first set of data was used to perform the OD, then the determined orbit was propagated tol the second set of data where the computed angular and range measurements are then compared with those really measured. It must be said that the aim of this study is not to achieve the most accurate orbit, but to ensure the re-observability of the object in the next pass.

First, an initial orbit determination based on only angular measurements is done to generate the apriori orbit, then this is improved via a least square adjustment using the measurements belonging to the first set of observations. For TOPEX, the first set of data correspond to the measurements acquired during one pass, and for the Glonass, during the first 12 minutes of observations. The OD procedure was repated, for each satellite, 3 times: the first processing used only the acquired angular measurements, the second processing only the ranges, and the third the merged measurements. Figure 9 and Figure 10 show the obtained residuals for TOPEX and the Glonass, respectively. The left plot shows the total angular residuals (sum of the Azimuth and Elevation residuals) while the right one shows those for the ranges. The total angular residuals represent a good discriminant for the re-observability of the object since if this error is bigger than half of the FoV of the instrument, the object is lost. As we have said in Paragraph 2, the FoV of the tracking camera is ~7 arcmin, therefore to reobserve an object the maximum tolerated error is 3.5 arcmin. Once the object is again in the FoV of the camera one can correct the pointing of the telescope and adjust the parameters which will allow the ranging to the satellite by the SLR system.

Focusing now on the results obtained for TOPEX (Figure 9), we can see how using only one observable we will not be able to observe the object during its next pass using the tracking camera. The obtained error in fact, is bigger than 2000 arcsec (>33 arcmin). The situation changes completely if we process merged measurements. In this case, the obtained error is less than 200 arcsec (3.3 arcmin) which is one order of magnitude better than before. Another interesting outcome is that the solution provided by the ranges, although their high accuracy, is worse than that obtained processing only angular measurements. This is not surprising and depends on the quality of the apriori orbit, on the short observation arc and on the fact that all 6 orbital elements are estimated in the OD process.

Looking at Figure 10, one can see that for the Glonass case we would be able to reobserve the object even using only ranges, this is essentially due to the fact that the set of data used for comparison is just ~20 minutes after the end of the first one. Nevertheless, the data fusion produces still an improvement of one order of magnitude in the accuracy of the estimated orbit. Looking at the range residuals we can see that in this case the only range solution is much closer to the merged one, which is several orders of magnitude better than the one obtained by the only angle solution. These are consequences of the type of observables (ranges and angles) and the length of the observation arc. The ranges provide information about the distance of the object which can be easily extrapolated during the same passage but they don't provide enough information to estimate correctly the orientation of the orbital plane. Vice versa, the angles allow the estimation of the orientation of the orbital plane but they do not provide direct measurements of the distance of the object. This produces a higher accuracy of the angles only orbit when looking at the angular residuals.

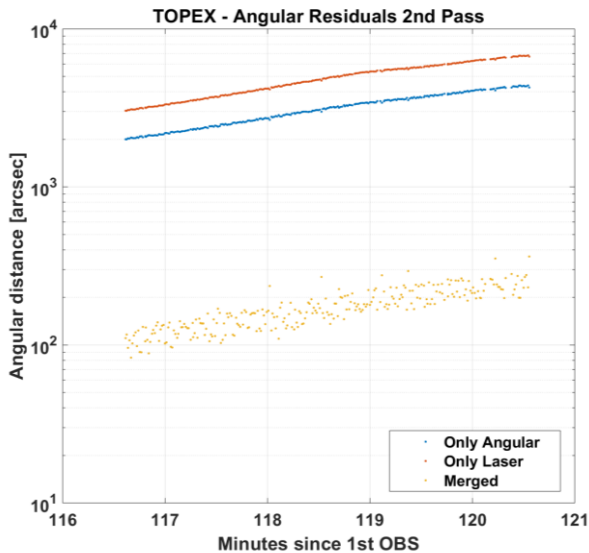 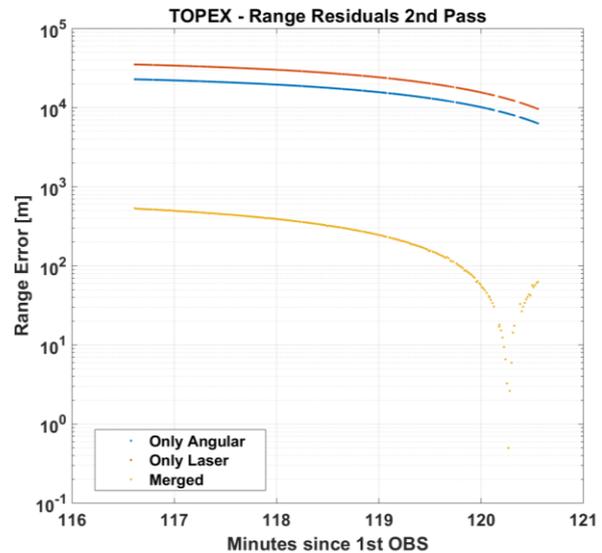

*Figure 9 TOPEX residuals at 2nd pass obtained propagating the OD results based on the observations acquired in the first pass.*

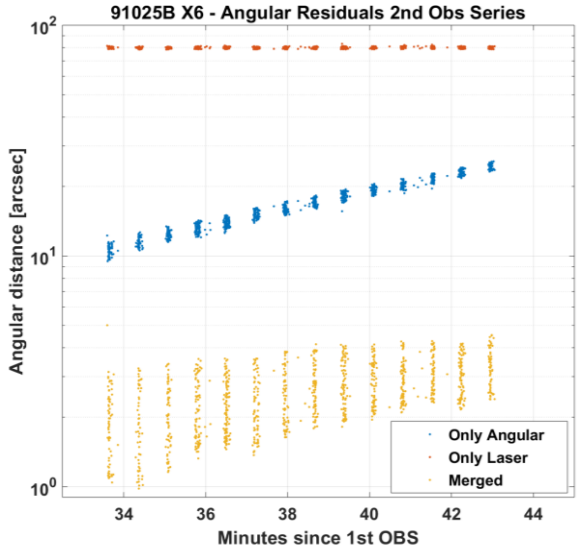 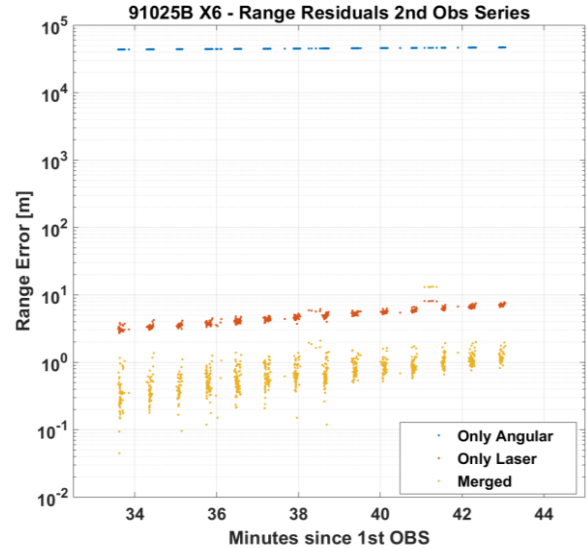

*Figure 10 Glonass residuals at the second portion of the pass obtained by propagating the OD results based on observations acquired in the first portion.*

## 6 CONCLUSION

In this paper we have shown the results obtained by the application of the night-tracking camera for space debris studies. The camera becomes necessary when we want to track space debris, characterized by ephemeris with poor accuracy, with the SLR system which has a relatively small FoV. The camera, correcting the pointing of the telescope in real time, allows us to track LEO and MEO defunct satellites with our SLR system. The main outcomes of the tracking are the angular positions of the object in the sky (Azimtuh and Elevation), its distance, and its brigthness. All these measurements are acquired synchronously with the timing accuracy provided by the SLR system. It must be said that the angular measurements are obtained directly from the encoder readings of the telescope without any astrometric data reduction process. This doesn't produce the most accurate angular measurements but at the same time we do not need to see stars in the image, which is a big advantage when observing fast satellites with a telescope with a relatively small FoV. Another important aspect is the measurement rate provided by the entire system (SLR and tracking camera) which provides 100 Hz range measurements and up to 30 Hz for the angular and brigthness data (when using full frame images).

We have shown how the acquired measurements can be used for both, the attitude and the orbit determination of space debris. The obtained results were validated through a comparison with database values and with real measurements. Particularly important are the outcomes obtained for the OD: the acquisition of both, angular and range measurements from one single pass, or just a small portion of it, ensures the reobsevability of the object during its next pass. This is also possible thanks to the quasi realtime pipeline developed to extract and process the data provided by the tracking camera.

The results, still preliminary, demonstrate the advantages and the possibilities connected to the usage of the tracking camera. Therefore, we will continue in the automatization process of the camera. Next steps to be performed will be the improvement of the laser pointing model on the camera, the automatic object recognition in the image and the consequent automatic correction of the telescope pointing. Then we will complete the automatization of the pipeline to process the measurements, perform an OD, generate and update the telescope ephemeris. The final goal of the project is the active tracking, in real time, of objects with a poorly known or unknown orbit.

One big limitation of the actual tracking camera is the kind of correction provided to the telescope. At the moment, azimuth and elevation corrections are provided; this recquires, for cases where the ephemeris accuracy is low, a continuous correction of the telescope which is possible only when the object reflects sunlight. Therefore, we lose the object when it enters shadow of the Earth; this could be prevented by changing the correction sent to the telescope from azimuth/elevation to along-, cross-track corrections. Finally, one last aspect which could be interesting to investigate is the application of the tracking camera during daytime observations.